\documentclass[12pt]{article}
\usepackage{cite}
\usepackage[a4paper,hscale=0.82,vscale=0.75]{geometry}
\usepackage[parfill]{parskip}
\usepackage{graphicx}
\usepackage{amssymb}

\usepackage{amsmath}
\usepackage{amsthm}
\usepackage{amscd}
\usepackage[all,cmtip]{xy} 
\usepackage{empheq}
\usepackage[british]{babel}
\usepackage[small,nohug]{diagrams} 
\title{Born-Rigid Flow and the AdS-CFT Correspondence\vspace{.8cm}}
\author{Ziyang Hu\footnote{\texttt{z.hu@damtp.cam.ac.uk}}\\
D.A.M.T.P.\\
 University of Cambridge}
\newcommand{\pd}{\partial}
\newcommand{\rs}{\mathbb{R}}

\newtheoremstyle{shape0}% name
  {9pt}%      Space above
  {9pt}%      Space below
  {}%         Body font
  {}%         Indent amount (empty = no indent, \parindent = para indent)
  {\bfseries}% Thm head font
  {.}%        Punctuation after thm head
  {.5em}%     Space after thm head: " " = normal interword space;
  {}%         Thm head spec (can be left empty, meaning `normal')

\newtheoremstyle{shape1}% name
  {9pt}%      Space above
  {9pt}%      Space below
  {\it}%         Body font
  {}%         Indent amount (empty = no indent, \parindent = para indent)
  {\bfseries}% Thm head font
  {.}%        Punctuation after thm head
  {.5em}%     Space after thm head: " " = normal interword space;
  {}%         Thm head spec (can be left empty, meaning `normal')

\newtheoremstyle{shape2}% name
  {9pt}%      Space above
  {9pt}%      Space below
  {}%         Body font
  {}%         Indent amount (empty = no indent, \parindent = para indent)
  {\itshape}% Thm head font
  {.}%        Punctuation after thm head
  {.5em}%     Space after thm head: " " = normal interword space;
  {}%         Thm head spec (can be left empty, meaning `normal')

\theoremstyle{shape1}
\newtheorem*{thm}{Theorem}
\newtheorem*{assump}{Assumption (to be proved)}
\newtheorem*{hnthm}{Herglotz-Noether theorem}
\theoremstyle{shape0}

\theoremstyle{shape2}
\theoremstyle{definition}

\begin{document}
\maketitle
\abstract{This paper reviews the concepts and assumptions of rigid flow in relativistic fluid mechanics, particularly the generalisation of the classical Herglotz-Noether theorem, that are relevant to the fluid approximation of the AdS-CFT dual of large rotating black-holes used by Bhattacharyya et al. We then give a brief outline of the recently found proof the generalised theorem.}

\section{Introduction}
\label{sec:introduction}

Recent work has attempted to predict universal features of large rotating black holes in AdS backgrounds through the AdS-CFT correspondence\cite{Bhattacharyya:2007p31}. The strategy is first to consider the dual description of black holes in terms of a quantum field theory on a conformal background, which is usually taken to be the Einstein static universe $S^{D-2}\times\text{time}$\cite{Maldacena:1997p9405, Witten:1998p9408}, and then pass to the thermodynamic limit of the quantum field theory, which was then argued to yield a \emph{classical} and \emph{dissipationless} relativistic fluid flowing on the conformal boundary of AdS. Physically, the most important cases are AdS$_{5}\times S^{5}$, AdS$_{7}\times S^{4}$ and AdS$_{4}\times S^{7}$ which arise from type IIB and M theory, but the general case is also interesting in its own right. Thus, assuming the use of duality and the passing to the limit are legitimate in the realm one is considering, understanding of the classical fluid immediately yields an understanding of the original black holes. 
Our aim in this paper is to make more rigorous the foundations for the calculations pertaining to the fluid flow part of the above-mentioned programme.  In particular, after some brief comments on the derivation of the fluid flow in the CFT case, we link the particular class of fluid flow arising in this case to the classical Born-rigid flow in special relativity and its generalisations, and then give a simple description of proofs of a few mathematical assumptions about fluid flow in conformally flat spacetime that are implicitly used in the calculations of the fluid description of AdS black holes, as given in \cite{Hu:2010p9410, Hu:2010p9409}.

\section{Fluid model in the AdS-CFT correspondence}
We will first review some key steps of the derivation of the fluid flow.
Let use first consider the problem of modelling a relativistic fluid flow: the flow is modelled as a normalised time-like vector field, that is,
\[
g_{\mu\nu}u^{\mu}u^{\nu}=-1
\]
where $g_{\mu\nu}$ is the metric of the conformal background spacetime $M$. This vector field is the set of the worldlines of the fluid particles. As in classical fluid mechanics, let us define a few convenient quantities. Define the induced metric with respect to the fluid as
\[
h^{\mu\nu}=g^{\mu\nu}+u^{\mu}u^{\nu}
\]i.e., $h^{\mu\nu}$ is the projection onto the tangent subspace orthogonal to the flow. Then we can decompose the spacelike components of $\nabla_{\mu}u_{\nu}$ (i.e., by projecting this quantity with $h^{\mu\nu}$) as the expansion
\[
\theta=u^{\mu}{}_{;\mu}
\]
the shear
\[
\sigma^{\mu\nu}=\frac{1}{2}(h^{\mu\sigma}u^{\nu}{}_{;\sigma}+h^{\nu\sigma}u^{\mu}{}_{;\sigma}-\frac{1}{n-1}\theta h^{\mu\nu})
\]
and the vorticity
\[
\omega^{\mu\nu}=\frac{1}{2}(h^{\mu}{}_{\sigma}h^{\nu}{}_{\lambda}u^{\sigma;\lambda}-h^{\mu}{}_{\sigma}h^{\nu}{}_{\lambda}u^{\lambda;\sigma})
\]
($n$ is the dimension of the spacetime under consideration). That this is really an orthogonal decomposition can be seen by writing out the stress tensor
\[
u_{\mu;\nu}-\dot u_{\mu}u_{\nu}=\omega_{\mu\nu}+\sigma_{\mu\nu}+\frac{1}{n-1}\theta
\]
where $\dot u^{\nu}=u^{\mu}{}_{;\sigma}u^{\sigma}$ is the acceleration. These quantities have physical interpretations suggested by their names; see \cite{ehlers1961}.

A flow described by such a vector field $u^{\mu}$ is completely general, and the key point in the analysis is that a flow on the CFT space given by the dual of a black hole must, to some first approximation, be \emph{dissipationless}, where as usual in the fluid mechanics case, being dissipationless means that the flow has no shear nor expansion, $\theta=0$ and $\sigma^{\mu\nu}=0$.
The no-shear condition is argued in a straightforward manner by recognising that in the perfect fluid approximation, we cannot have entropy production and a non-zero $\sigma^{\mu\nu}$ produces entropy. The expansion-free condition $\theta=0$, on the other hand, is geometrical in nature. It should be recognised that in this paper and in \cite{Bhattacharyya:2007p31} we are actually calculating quantities of conformal geometry using the language of Riemannian geometry, and in conformal geometry proper, the trace of the stress tensor is not physical at all.  This is perhaps most clearly seen by recognising, as in \cite{Bhattacharyya:2007p31}, that the trace is related to the (Riemannian) scalar curvature of the background space, but the scalar and Ricci part of the curvature can be set to arbitrary value by applying a conformal transformation to the space. Hence, assuming that such a conformal transformation has already been applied, we can safely assume that $\theta=0$. There is a caveat, however: even if we started with a particular Riemannian model of a conformal space, say $S^{3}\times\rs$, after setting $\theta=0$ we can no longer safely working using the usual Riemannian metric on $S^{3}\times\rs$, otherwise inconsistencies might occur. Therefore, any geometrical results that we need to apply in such a situation has to be results valid for \emph{all} conformally flat spacetime; validity for $S^{3}\times\rs$ alone is insufficient.

\section{Relation to the classical Herglotz-Noether theorem}
\label{sec:born-rigid-flow}
Now we are satisfied that in the fluid approximation of the dual description of black holes, the relevant fluid is a dissipationless (shear-free and expansion-free) fluid flowing in a conformally flat spacetime. The following assumption further simplifies the matter so that concrete calculations can now be carried out: 
\begin{assump}
  {In a conformally flat spacetime, the only dissipationless and rotating ($\omega{^{\mu\nu}}\neq 0$) fluid flow are those flow that coincide with Killing flows (isometric flows).}
\end{assump}
If we can turn this assumption into a theorem, then the calculations in the dual description of black holes are certainly given a more rigorous footing. The place to start is to recognise that if we replace ``conformally flat'' by simply ``flat'', and restrict considerations to $3+1$ Minkowski spacetime, then this is the case where the classical Herglotz-Noether theorem applies:
\begin{hnthm}
  In $4$-dimensional Minkowski spacetime, rotating Born-rigid flow are isometric.
\end{hnthm}
Here Born-rigid has the same meaning as dissipationless that we have been using. That dissipationless flows can be called ``rigid'' is seen by the equivalence of shear-free and expansion-free conditions and
\begin{equation}
  \label{eq:1}  
\mathcal{L}_{u}h^{\mu\nu}=0,
\end{equation}
i.e., the orthogonal distance along the fluid flows are preserved (the derivation is not difficult and can be found in, e.g., \cite{pirani1964}). This definition, originally formulated for flat spacetime, remains valid in curved spacetime.  Another thing to keep in mind is that, by definition, our flow vector field $u^{\mu}$ is normalised, whereas a Killing vector field of course cannot be meaningfully normalised. Recall that a vector field $V^{\mu}$ is Killing if and only if $V_{(\mu;\nu)}=0$, hence our flow is isometric if and only if
\begin{equation}
  \label{eq:2}  
(\lambda u_{\mu})_{;\nu}+(\lambda u_{\nu})_{;\mu}=0.
\end{equation}
Where $\lambda$ is \emph{any} positive scalar function defined on the spacetime. What the Herglotz-Noether theorem says is that under the assumptions of the theorem, \eqref{eq:1} and \eqref{eq:2} are the same condition (the equations are ``integrable'').

The Herglotz-Noether theorem is rather old, discovered in 1910 independently by Herglotz \cite{herglotz1910} and Noether \cite{noether1910}. Pirani and Williams \cite{pirani1962} in 1962 gives the theorem firmer foundation by discussing the ``integrability conditions'' for such a rigid flow, which also makes possible a new, better proof for the theorem using tensor calculus. Indeed, it is the proof by Pirani and Williams, instead of the original proofs by Herglotz and Noether, that is subsequently followed on in the literature. The Pirani-Williams proof was reproduced by Pirani more succinctly in the lectures in \cite{pirani1964} which, in addition,  gives a counterexample showing that in the general Riemannian spacetime of general relativity (4-dimensional), the theorem is false. All these are summarised and reviewed by Giullini in the 2006 review \cite{Giulini:2006p66}, using more modern notation. All the above investigations are restricted to $(3+1)$ dimensions and are only concerned with the flat case, apart from the counterexample given in the general case. A different line of development by Wahlquist and Estabrook from 1964 to 1967 \cite{Estabrook:1964p2608,Wahlquist:1966p2548,Wahlquist:1967p2556} used dyadic analysis which, by the choice of method, again restricted to $(3+1)$ dimensions. Due to the hermetic language of dyadic analysis, the this approach was not followed subsequently by others.

\section{Introduction to the proof}
\label{sec:introduction-proof}

Even though we really need a generalisation of the theorem, it is helpful to ponder the proof of the more straightforward theorem to give us some hint on the possible paths of generalisations.  The details of the proof in \cite{pirani1962} is not very easy to follow since one is forced to do very messy tensor index manipulations, but the basic idea is clear. After some fooling around with brute-force calculation on the equivalence of conditions \eqref{eq:1} and \eqref{eq:2} (the integrability condition), one is forced to conclude that a more clever approach is needed. The problem itself is more or less well-defined (the equivalence of two conditions for a fluid flow), however neither of these two conditions and their associated equations can be said to be in ``normal'' form and comparison is difficult. Therefore, the obvious strategy is to put the problem into more ``normal'' form, i.e., by reduction of free variables. It should be observed that \eqref{eq:1} implies that, at least locally, the induced metric really behaves as a Riemannian metric for \emph{some} quotient space. Then, if one can use a coordinate system were the horizontal part of the coordinates are constrained by the induced metric only, one has achieved the goal of a reduction. Indeed this is the approach taken in \cite{pirani1962}: we take a coordinate system $x^{\mu}$ where 
\[
u^{\mu}=(1,0,0,0),\qquad\frac{\pd h_{\mu\nu}}{\pd x^{0}}=0,
\]
note that the condition uses \emph{partial} derivative and hence are coordinate-dependent. Also, the \emph{existence} of such a coordinate system is non-trivial to justify. Simplification occurs since in this coordinate system,
\[
\mathcal{L}_{u}T_{\mu\nu\dots}=\frac{\pd}{\pd x^{0}}T_{\mu\nu\dots}.
\]
In such a set-up, we have the quantity $h_{\mu\nu}$ depending only on its spacelike coordinates $h_{ij}$. Therefore we can introduce two connections on the total spacetime we are considering: one connection arises from the metric $g_{\mu\nu}$ and is concerned with parallelism in the total space, and the other connection arises from the metric $h_{ij}$ and is only concerned with \emph{horizontal} movement. The key step is to translate as many formulae (especially those for the vorticity, shear and expansion) from those involving $g_{\mu\nu}$ to those involving $h_{ij}$ and $\pd/\pd x^{0}$ only. After this is done, we are left with a set of much simpler (but still not so simple) equations, and the requirement that the total space be flat and a few further tricks will manipulate the conditions \eqref{eq:1} and \eqref{eq:2} to be the same condition. The rather strange requirement that the fluid has to be rotational for the theorem to hold occurs because the non-vanishing of the vorticity is necessary to make one key equation non-trivial to place further constraints on the coordinates. This approach is only valid in ($3+1$) dimensions because in the manipulation the duality of rank-2 antisymmetric tensor and a vector in the spacelike subspace is used to write the vorticity as a vector, which is impossible to do in higher dimensions.
 It is clear that to further generalise the theorem, we need to devise an approach that does not rely on the duality of tensors and vectors in 3-dimensions.

It is important to note is that the relations between the two metrics $h_{ij}$ and $g_{\mu\nu}$ are subtle, that is, their Christoffel symbols $\Gamma^{\mu}{}_{\nu\lambda}$ and $\tilde\Gamma^{i}{}_{jk}$, even when the indices are the same, are in general not equal. A geometrical way to say this is that the ``reduced space'' in which $h_{ij}$ lives is not an integrable surface in the total space.

Based on the observations on the proof of the classical theorem made above, in our proof of the generalised theorem, we will attempt the following:
\begin{itemize}
\item Before doing \emph{any} calculations at all, we will work abstractly and reduce the space where the problem lives to a much smaller one.
\item We will avoid using coordinates at all costs. Indeed, we will not use any coordinate system on the base space at all.
\end{itemize}
The proof of the generalised theorem can be found in \cite{Hu:2010p9409}. Here we give a conceptual summary. In the proof of the classical theorem, much progress was made based on a careful use of special coordinates and the introduction of the induced metric and connections. However, the use of tensor equations make the problem over-complicated since often ad-hoc manipulations and tricks are needed, and it is awkward to discuss the symmetries processed by the special coordinates. We can overcome both of these problems by using a \emph{coframing} of the manifold. Indeed, let $\theta^{\mu}$ be a set of linearly independent, orthonormal 1-forms on the manifold, with $\theta^{0}$ dual to the vector $u^{\mu}$. Here it must be remembered that $\theta^{0}$ is one single 1-form on the manifold, hence it has meaning independent of coordinates, and is dual to $u^{\mu}$. By contrast, $u^{0}$ has no coordinate independent meaning. Hence if we only deal with $\theta^{\mu}$ and their exterior derivatives, we are free from any special coordinates. Note that the symmetry of the problem is clear: the group $SO(1,n-1,\rs)$ acts on $\theta^{\mu}$ via the defining representation.

The exterior derivatives are
\[
d\theta^{\mu}=\theta^{\nu}\wedge\omega^{\mu}{}_{\nu}
\]
the 1-forms $\omega^{\mu}{}_{\nu}$ are another set of 1-forms. Since the $\theta^{\mu}$ are a coframing, the $\omega^{\mu}{}_{\nu}$ can be written as linear combinations of the former. However, to really facilitate calculations it is helpful to consider as our space not $M$ on which the flow occurs, but (locally, at least) $M\times G$ where $G$ is the symmetry group $SO(1,n-1,\rs)$, i.e., considering the principal bundle suitable for this problem. This is Cartan's approach to geometry. Then, $\theta^{\mu}$ and $\omega^{\mu}$ are linearly independent on this larger space, and they span the cotangent space. Differentiating again we obtain the curvature (on the principal bundle):
\[
d\omega^{\mu}{}_{\nu}=\omega^{\lambda}{}_{\nu}\wedge\omega^{\mu}{}_{\lambda}+\Omega^{\mu}{}_{\nu}
\]
where the curvature $\Omega^{\mu}{}_{\nu}$ is a 2-form on the bundle. The condition of flatness can be expressed as
\[
\Omega^{\mu}{}_{\nu}=0
\]
whereas the condition for conformal flatness says the Weyl subspace of $\Omega^{\mu}{}_{\nu}$ vanishes. These expressions are in general much simpler than those in terms of metric and Christoffel symbols and will be our constraints on the problem.

Since we have \emph{already} chosen uniquely $\theta^{0}$, we really should have a reduction of the principal bundle, since the structure group is now only $SO(n-1,\rs)$, acting on $\theta^{i}$ only. In the reduced bundle, the forms $\theta^{\mu}$, $\omega^{\mu}{}_{\nu}$ still span, however they are no longer linearly independent. Indeed, the redundant ones can be taken to be
\[
\omega^{0}{}_{i}=\omega^{i}{}_{0}=K_{i}\omega^{0}+M_{ij}\omega^{j}
\]
where $K_{i}$ and $M_{ij}$ are sets of \emph{scalar functions} on the principal bundle and remember that all indices are the so-called \emph{tangent space indices}, i.e., they label different quantities, not components of the same quantity.

Up until now we have only been formulating the old tensor proof into a new language. However, the following step, which is crucial to the proof of the generalised theorem, is impossible in the tensor language: we recognise that \emph{if} we align the $\theta^{i}$ properly along each flow worldline, then the principal bundle is further reduced: on each flow worldline we have \emph{no} degree of freedom in $G$ left, though we still have $SO(n-1,\rs)$ when we jump from worldline to worldline. In the tensor language, since the forms $\theta^{i}$ are \emph{not} integrable in general, there does not exist special coordinates in which $\theta^{i}=dx^{i}$, and hence it is not possible to talk about this alignment and subsequent reduction of structure group.

After this reduction, and using the fact that the flow is \emph{rigid}, the forms $\theta^{i}$ (without $\theta^{0}$) are actually well-defined 1-forms on the quotient space $M/\{\text{worldlines of the flow}\}$. Hence we can repeat the ``lifting onto the principal bundles'' approach and obtain
\[
d\theta^{i}=\theta^{j}\wedge\widetilde\omega^{i}{}_{j},\qquad d\widetilde\omega^{i}{}_{j}=\widetilde\omega^{k}{}_{j}\wedge\widetilde\omega^{i}{}_{k}+\widetilde\Omega^{i}{}_{j}
\]
and there is the relation
\[
\widetilde\omega^{i}{}_{j}=\omega^{i}{}_{j}-M^{i}{}_{j}\omega^{0}.
\]
Now it can be proved that $M_{ij}=0$ when $i=j$ since the decomposition of $\omega^{0}{}_{i}$ has to be compatible with the Lie algebra of the group. Also, $M_{ij}$ and $K_{i}$ corresponds exactly to vorticity and acceleration, and, miraculously, expansion and shear does not make any appearance at all (not surprisingly, this is due to the requirement that $\widetilde\omega^{i}{}_{j}$ exists, which is the same as saying that the flow is rigid). \emph{A priori}, the ``antisymmetry'' (we need to keep in mind that this is an antisymmetry among labelled functions, not the antisymmetry of a tensor) of $M_{ij}$ is the only constraint given by the rigidity of the flow; any other constraint must come from the geometry (curvature) of the background space. By contrast, the rotational isometry condition gives the following constraint in addition to the antisymmetry of $M_{ij}$:
\begin{equation}
  \label{eq:3}
  \dot M_{ij}=0
\end{equation}
where we made the further simplification by \emph{simulating} the usual covariant derivatives by expanding exterior derivatives of scalars on the principal bundles as:
\[
df = \dot f \theta^{0}+f_{;i}{\theta^{i}}+\text{terms linear in $\widetilde\omega^{i}{}_{j}$}.
\]

Actually, for the economy of notation, we have been using the same symbols for related quantities living on different spaces. A more detailed picture is the following:
\[
\xymatrix{
M\times G_{1}&\ar[l]M\times G_{2}&\ar[l]M\times G_{3}\ar[r]& M/\{\text{worldlines}\}\times G_{4}}
\]
where
\[
G_{1}=SO(1,n-1,\rs),\qquad G_{2}=G_{3}=G_{4}=SO(n-1,\rs)
\]
but $G_{2}$ and $G_{4}$ acts on the forms in the usual way, whereas $G_{3}$ only has effect when we move from worldline to worldline and once we fix an element along a point in a worldline we have fixed it on all of the worldline. Also note the directions of the arrows: all equalities occur in the space $M\times G_{3}$, where forms are pulled back by the appropriate maps. Hence, for example, $\omega^{\mu}{}_{\nu}$ really lives on $M\times G_{1}$, whereas $\widetilde\omega^{i}{}_{j}$ really lives on $M/\{\text{worldlines}\}\times G_{4}$.

After all these set up, we are ready to begin our calculations. All calculations will occur on $M\times G_{3}$. 
 We find that our constraints that the space is flat or conformally flat can be written as a set of second order polynomial equations involving only the curvature functions for the space $M$ and $M/\{\text{worldlines}\}$ (some of which we will set to zero) and the functions (which can be taken to be indeterminates)
\[
M_{ij},\quad \dot M_{ij},\quad M_{ij;k}, \quad K_{i},\quad \dot K_{i},\quad K_{i;j}.
\]
These polynomial equations can then be manipulated to give the condition for isometric (Killing) flow \eqref{eq:3}, which proves the theorem.

\section{Further comments}
\label{sec:furth-comm-concl}

We have mentioned before that this particular case of Born-rigid flow in a conformally flat spacetime amounts to doing conformal geometry using the language of Riemannian geometry. Hence, it is no surprise that, our already proven generalised theorem is equivalent to the following \cite{Hu:2010p9409}
\begin{thm}
  A conformally rigid flow is conformally isometric in a flat conformal geometry if the flow has non-vanishing vorticity.
\end{thm}
Here conformally isometry means the flow changes the representative conformal metric up to scale only, conformal flatness means the usual thing in conformal geometry, and conformal rigidity of flow is a flow that, in a representative Riemannian model, is shear-free but no necessarily expansion-free (as expansion can be ``scaled away'' in a conformal geometry anyway). The equivalence of this theorem and the theorem in the Riemannian language shows that these changes of the conditions and conclusions ``cancel'' each other.

These theorems facilitate calculations in that they are actually \emph{uniqueness theorems}: if we are dealing with rotational rigid flows and we can find some Killing flows that satisfy the conditions, we know there are no others.

\nocite{gibbonspri}
\nocite{born1909}
\nocite{herglotz1910}
\nocite{noether1910}
\nocite{pirani1962}
\nocite{rayner1959}
\nocite{BONeill:1966p7043}
\nocite{ehlers1961}
\nocite{synge1960}
\nocite{molino1988}
\nocite{Boyer:1965p2926}
\nocite{Robinson:1983p3530}
\nocite{tensor}
\bibliography{pub2}{}

\begin{thebibliography}{10}

\bibitem{Bhattacharyya:2007p31}
Sayantani Bhattacharyya, Subhaneil Lahiri, R~Loganayagam, and Shiraz Minwalla.
\newblock Large rotating {AdS} black holes from fluid mechanics.
\newblock {\em arXiv}, Aug 2007, 0708.1770v2.

\bibitem{born1909}
M~Born.
\newblock {\em Ann.~der Physik}, 30, 1909.

\bibitem{Boyer:1965p2926}
R~Boyer.
\newblock Rigid frames in general relativity.
\newblock {\em Proceedings of the Royal Society of London. Series A}, Jan 1965.

\bibitem{Estabrook:1964p2608}
F~Estabrook and H~Wahlquist.
\newblock Dyadic analysis of space-time congruences.
\newblock {\em Journal of Mathematical Physics}, Jan 1964.

\bibitem{gibbonspri}
Gary Gibbons.
\newblock Private communication.

\bibitem{Giulini:2006p66}
Domenico Giulini.
\newblock Algebraic and geometric structures of special relativity.
\newblock {\em arXiv}, Feb 2006, math-ph/0602018v2.

\bibitem{herglotz1910}
G~Herglotz.
\newblock {\em Ann.~der Physik}, 31, 1910.

\bibitem{Hu:2010p9410}
Ziyang Hu.
\newblock A modern view of the classical herglotz-noether theorem.
\newblock {\em arXiv}, math-ph, Apr 2010, 1004.1935v3.

\bibitem{Hu:2010p9409}
Ziyang Hu.
\newblock On the general problem of structure-preserving submersions.
\newblock {\em arXiv}, math-ph, Aug 2010, 1008.4084v1.

\bibitem{ehlers1961}
P~Jordan, J~Ehlers, W~Kundt, and R~K Sachs.
\newblock {\em Acad.~Wiss.~Lit., Mainz, Abh.~Math-naturw., Kl.}, 791, 1961.

\bibitem{Maldacena:1997p9405}
Juan~M Maldacena.
\newblock The large n limit of superconformal field theories and supergravity.
\newblock {\em arXiv}, hep-th, Nov 1997, hep-th/9711200v3.

\bibitem{molino1988}
P~Molino.
\newblock {\em Riemannian foliations}.
\newblock Birkhauser, Boston, 1988.

\bibitem{noether1910}
F~Noether.
\newblock {\em Ann.~der Physik}, 31, 1910.

\bibitem{BONeill:1966p7043}
B~O'Neill.
\newblock The fundamental equations of a submersion.
\newblock {\em Michigan Math. J}, Jan 1966.

\bibitem{pirani1962}
F~A~E Pirani and G~Williams.
\newblock Rigid motion in a gravitational field.
\newblock {\em S\'e{}minaire Janet}, 5, 1962.

\bibitem{rayner1959}
C~B Rayner.
\newblock {\em C.~R.~Acad.~Sci Paris}, 248, 1959.

\bibitem{Robinson:1983p3530}
I~Robinson and A~Trautman.
\newblock Conformal geometry of flows in n dimensions.
\newblock {\em Journal of Mathematical Physics}, Jan 1983.

\bibitem{synge1960}
J~L Synge.
\newblock {\em Relativity: the special theory}.
\newblock North-Holland Publ.~Co., Amsterdam, 1960.

\bibitem{tensor}
A~H Thompson.
\newblock The conformal generalisation of the {Herglotz-Noether} theorem.
\newblock {\em Tensor, N.S.}, 19, 1968.

\bibitem{pirani1964}
A~Trautman, F~A~E Pirani, and H~Bondi.
\newblock {\em Lectures on general relativity}.
\newblock Prentice-Hall, 1964.

\bibitem{Wahlquist:1966p2548}
H~Wahlquist and F~Estabrook.
\newblock Rigid motions in {E}instein spaces.
\newblock {\em Journal of Mathematical Physics}, Jan 1966.

\bibitem{Wahlquist:1967p2556}
H~Wahlquist and F~Estabrook.
\newblock Herglotz-{N}oether theorem in conformal space‐time.
\newblock {\em Journal of Mathematical Physics}, Apr 1967.

\bibitem{Witten:1998p9408}
Edward Witten.
\newblock Anti de sitter space and holography.
\newblock {\em arXiv}, hep-th, Feb 1998, hep-th/9802150v2.

\end{thebibliography}
\bibliographystyle{hplain}
\addcontentsline{toc}{section}{References}
\end{document}